# Representing higher-order dependencies in networks

Jian Xu,[1,3] Thanuka L. Wickramarathne,[2,3,4] Nitesh V. Chawla[1,3,4]*

[1]Department of Computer Science and Engineering, University of Notre Dame, USA

[2]Department of Electrical Engineering, University of Notre Dame, USA

[3]Interdisciplinary Center for Network Science and Applications, University of Notre Dame, USA

[4]Environmental Change Initiative, University of Notre Dame, USA

*To whom correspondence should be addressed; E-mail: nchawla@nd.edu

**Abstract**
To ensure the correctness of network analysis methods, the network (as the input) has to be a sufficiently accurate representation of the underlying data. However, when representing sequential data from complex systems such as global shipping traffic or web clickstream traffic as networks, conventional network representations that implicitly assume the Markov property (first-order dependency) can quickly become limiting. This assumption holds that when movements are simulated on the network, the next movement depends only on the current node, discounting the fact that the movement may depend on several previous steps. However, we show that data derived from many complex systems can show up to fifth-order dependencies. In these cases, the oversimplifying assumption of the first-order network representation can lead to inaccurate network analysis results. To address this problem, we propose the Higher-Order Network (HON) representation that can discover and embed variable orders of dependencies in a network representation. Through a comprehensive empirical evaluation and analysis, we establish several desirable characteristics of HON, including accuracy, scalability, and direct compatibility with the existing suite of network analysis methods. We illustrate how HON can be applied to a broad variety of tasks, such as random walking, clustering, and ranking, and we demonstrate that by using it as input, HON yields more accurate results without any modification to these tasks.

## Introduction

Today's systems are inherently complex, whether it is the billions of people on Facebook powering a global social network, the transportation networks powering the commute and the economy, or the interacting neurons powering the coherent activity in the brain. Complex systems such as these are made up of a number of interacting components that influence each other, and network-based representation has quickly emerged as the norm by which we represent the rich interactions among the components of such a complex system. These components are represented as nodes in the network, and the edges or links between these nodes represent the (ranges and strengths of) interactions. This conceptualization raises a fundamental question: *given the data, how should one construct the network representation such that it appropriately captures the interactions among the components of a complex system?*

A common practice to construct the network from data (in a complex system) is to directly take the sum of pairwise connections in the sequential data as the edge weights in the network—e.g., the sum of traffic between locations in an interval *(1–4)*, the sum of user traffic between two web pages, and so on

*(5–7)*. This direct conversion, however, implicitly assumes the Markov property *(8)* (first-order dependency) and loses important information about dependencies in the raw data. For example, consider the shipping traffic network among ports, where the nodes are ports and the edges are a function of the pairwise shipping traffic between two ports. When interactions are simulated on the network, such as how the introduction of invasive species to ports is driven by the movements of ships via ballast water exchange, the next interaction (port-port species introduction) only depends on the current node (which port the ship is coming from), while in fact the interaction may be heavily influenced by the sequence of previous nodes (which ports the ship has visited before). Another example is user clickstreams on the web, where nodes are web pages and interactions are users navigating from one web page to another. A user's next page visit not only depends on the last page but is also influenced by the sequence of prior clicks. Thus, there are higher-order dependencies in networks and not just the first-order (Markovian) dependency as captured in the common network representation. In this paper, we focus on deriving the network based on the specific set of interactions, namely the interactions induced by movements among components of a complex system, wherein the sequence of movement patterns becomes pivotal in defining the interactions.

Let us again consider the process of constructing a network from the global shipping complex system by incorporating the movements from the ship trajectories *(9, 10)* (Fig. 1A). Conventionally *(1–7, 9)*, a network is built by taking the number of trips between port pairs as edge weights (Fig. 1B). When ship movements are simulated on this first-order network, according to the network structure where the edge *Singapore → Los Angeles* and the edge *Singapore → Seattle* have similar edge weights, a ship currently at Singapore has similar probabilities of going to Los Angeles or Seattle, no matter how it arrived at Singapore. In reality, the global shipping data indicate that a ship's previous stops before arriving at Singapore do influence the ship's next movement: the ship is more likely to continue on to Los Angeles if it came from Shanghai, and more likely to go to Seattle if it came from Tokyo. A first-order network representation fails to capture important information like this because in every step, the flow of traffic on the network is simply aggregated and mixed. As a consequence, trajectories simulated on the first-order network do not follow true ship movement patterns. By contrast, the higher-order network, by breaking down the node *Singapore* into *Singapore|Tokyo* and *Singapore|Shanghai* (Fig. 1C), can better guide the movements simulated on the network. As ships can translocate species along intermediate stops via partial ballast water exchanges *(11)*, the ability to distinguish between these cases is critical for producing accurate species introduction probabilities for each port.

Such higher-order dependencies exist ubiquitously and are indispensable for modeling vehicle and human movements *(12)*, email correspondence, article and web browsing *(13–15)*, conversations *(16)*, stock market *(17)*, and so on. While higher-order dependencies have been studied in the field of time series *(17, 18)*, information theory *(19)*, frequent pattern mining *(20)*, next location prediction *(21)*, variable order Markov *(22–24)*, hidden Markov model *(25)*, Markov order estimation *(26–28)*, they have focused on the stochastic process, rather than on how to represent higher-order dependencies in networks to adequately capture the intricate interactions in complex systems. In the field of network science, the frontier of addressing the higher-order dependencies still remains at the stage of assuming a fixed second order of dependency when constructing the network *(12, 29–32)* or using multiplex networks *(33)*, and there is neither a thorough discussion beyond second-order dependencies, nor a systematic way of representing dependencies of variable orders in networks. While there have also been efforts to incorporate higher-order network structures for clustering *(34, 35)*, ranking *(36)*, and so on, these approaches need to modify existing algorithms and are application-specific. As a result, these methods are not generalizable to broader applications, while we expect a network representation that is agnostic to the end-analysis methods (more discussions in Materials and Methods).

In this paper, we present a novel and generalizable process for extracting higher-order dependencies in the sequential data and constructing the Higher-Order Network (HON) that can represent dependencies of variable orders derived from the raw data. We demonstrate that HON is: (a) more reflective of the underlying real-world phenomena (for example, when using HON instead of a first-order network to represent the global shipping data, the accuracy is doubled when simulating a ship's next movement on the network, and is higher by one magnitude when simulating three steps); (b) efficient in scaling to higher orders, since auxiliary higher-order nodes and edges are added to a first-order network only where necessary; and (c) consistent with the conventional network representation, allowing for a variety of existing network analysis methods and algorithms to run on HON without modification. These algorithms and methods produce considerably different and more accurate results on HON than on a first-order network, thus demonstrating the broad applications and potential influences of this novel network representation.

We analyze a variety of real-world data including global shipping transportation, clickstream web browsing trajectories, and Weibo retweet information diffusions. We show that some of them have dependencies up to the fifth order, which the conventional first-order network representations or the fixed second-order network representations simply cannot capture, rendering the downstream network analyses tools such as clustering and ranking with limited and possibly erroneous information about the actual interactions in data. We also validate HON's ability to reveal higher-order dependencies on a synthetic data set, where we introduced dependencies of variable orders through a process completely independent of the construction of HON. We show that HON accurately identifies all the higher-order dependencies introduced.

## Results

We start with an examination of the conventional network representation, showing its limitations and formally introducing the higher-order network. Then with multiple real-world and synthetic data sets, we compare our proposed network representation with the conventional ones in terms of accuracy, scalability, and observations drawn from network analysis tools.

### The higher-order network (HON) representation

Conventionally, a network (also referred to as a graph) $G=(V, E)$ is represented with vertices or nodes $V$ as entities (e.g., places, web pages, etc.) and edges or links $E$ as connections between pairs of nodes (e.g., traffic between cities, user traffic between web pages, etc.). Edge weight $W(i \rightarrow j)$ is a number associated with an edge $i \rightarrow j$ representing the intensity of the connection, which is usually assigned as the (possibly weighted) sum of pairwise connections $i \rightarrow j$ (e.g. the daily traffic from $i$ to $j$) in data *(1–7, 9)*.

A wide range of network analysis methods such as PageRank for ranking *(37)*, MapEquation *(38)* and walktrap *(39)* for clustering, and link prediction methods *(40, 41)* use random walking to simulate movements on networks (e.g. ships traveling between ports, users clicking through web pages, etc.). If the location of a random walker at time $t$ is denoted as a random variable $X_t$ where $X$ can take values from the node set $V$, then, conventionally *(40, 42)*, the transition probability from node $i_t$ to the next step $i_{t+1}$ is proportional to the edge weight $W(i \rightarrow i_{t+1})$:

$$P(X_{t+1} = i_{t+1} \mid X_t = i_t) = \frac{W(i_t \rightarrow i_{t+1})}{\sum_j W(i_t \rightarrow j)} \tag{1}$$

This Markovian nature of random walking dictates that every movement simulated on the network is only dependent on the current node. In the conventional first-order network representation, every node maps to a unique entity or system component, so that every movement of a random walker is only dependent on a single entity (in Fig. 1B is Singapore). Data with higher-order dependencies that involve more than two entities, such as "ships coming from Shanghai to Singapore are more likely to go to Los Angeles" in the global shipping data, cannot be modeled via the conventional first-order network representation. Thus, the simulation of movement performed on such networks will also fail to capture these higher-order patterns.

To represent higher-order dependencies in a network, we need to rethink the building blocks of a network: nodes and edges. Instead of using a node to represent a single entity (such as a port in the global shipping network), we break down the node into different higher-order nodes that carry different dependency relationships, where each node can now represent a series of entities. For example, in Fig. 1C, Singapore is broken down into two nodes, Singapore *given* Tokyo as the previous step (represented as *Singapore|Tokyo*), and Singapore *given* Shanghai as the previous step (represented as *Singapore|Shanghai*). Consequently, the edges *Singapore|Shanghai* → *Los Angeles* and *Singapore|Shanghai* → *Seattle* can now involve three different ports as entities and carry different weights, thus representing second-order dependencies. As the out-edges here are in the form of $i|h \rightarrow j$ instead of $i \rightarrow j$, a random walker's transition probability from node $i|h$ to node $j$ is:

$$P(X_{t+1} = j \mid X_t = (i \mid h)) = \frac{W(i \mid h \rightarrow j)}{\sum_k W(i \mid h \rightarrow k)} \tag{2}$$

so that although a random walker's movement depends only on the current node, it now depends on multiple entities in the new network representation (as in Fig. 1C), thus being able to simulate higher-order movement patterns in the data. More importantly, this new representation is consistent with conventional networks and compatible with existing network analysis methods, because the data structure of HON is the same as the conventional network (the only change is the labeling of nodes). This makes it easy to use HON instead of the conventional first-order network as the input for network analysis methods, with no need to change the existing algorithms. The algorithm to construct HON is in Materials and Methods.

Rosvall et al. *(12)* consider a higher-order dependency, albeit with a fixed second-order assumption. They propose a network representation comprised of "physical nodes" and "memory nodes". As we will show with experiments, variable orders of dependencies can co-exist in the same data set, and can be up to the fifth order in our data. So if the dependency is assumed as fixed second order, it could be redundant when first-order dependencies are sufficient, and could be insufficient when higher-order dependencies exist. In HON, every node can represent an arbitrary order of dependency, so variable orders of dependencies can co-exist in the same network representation, as shown in Fig. 1D. For example, the fourth-order dependency relationship following the path of *Tianjin → Busan → Tokyo → Singapore* can now be represented as a fourth-order node *Singapore|Tokyo,Busan,Tianjin*; the second-order path *Shanghai → Singapore* is now a node *Singapore|Shanghai*; first-order relationships are now in a node *Singapore|.* Yet these nodes of variable orders all represent the same physical location Singapore. Compared with fixed order networks, we will show that our representation is compact in size by using variable orders and embedding higher-order dependencies only where necessary.

While the hypergraph *(43)* looks similar to HON in that its edges can connect to multiple nodes at the same time, it cannot directly represent dependencies. The reason is that dependencies are ordered relationships, but in a hypergraph the nodes connected by hyperedges are unordered, e.g., in the shipping example, an edge in a hypergraph may have the form of set{*Tokyo,Busan,Tianjin*} → set{*Singapore*}, where Tokyo, Busan, and Tianjin are interchangeable and cannot represent the path of the ship before arriving at Singapore. On the contrary, the edges in HON have the form of {*Tokyo|Busan,Tianjin*} → {*Singapore|Tokyo,Busan,Tianjin*} where the entities in nodes are not interchangeable. Thus HON can represent dependencies of arbitrary order.

**Higher-order dependencies in data revealed by HON**

First we show that HON can correctly extract higher-order dependencies from synthetic data. The synthetic data set has 10,000,000 generated movements, based on the pre-defined 10 second-order dependencies, 10 third-order dependencies, and 10 fourth-order dependencies (see Supplementary Material Note 1 for details). On this synthetic data with known variable orders of dependencies, HON (a) correctly captures all 30 of the higher-order dependencies out of the 400 first-order dependencies, with variable orders (from second-order to fourth-order) of dependencies mixed in the same data set correctly identified; (b) does not extract false dependencies beyond the fourth order even if a maximum order of five is allowed; (c) determines that all other dependencies are first-order, which reflects the fact that there is no other higher-order dependency in the data.

We then explore higher-order dependencies in real data: the global shipping data containing ship trajectories among ports, the clickstream data containing user browsing trajectories among web pages, and the retweet data containing information diffusion paths among users (see Supplementary Material Note 1 for details). The global shipping data reveals variable orders of dependencies up to the fifth order, indicating that a ship's movement can depend on up to five previous ports it has visited. The clickstream data also shows variable orders of dependencies up to the third order, indicating that the page a user will visit can depend on up to three pages the user has visited before, matching the observation in another study on web user browsing behaviors *(14)*. The fact that dependencies of variable orders up to the fifth order exist in real data further justifies our approach of representing variable and higher-order dependencies instead of imposing a fixed first or second order. On the contrary, the retweet data (recording information diffusion) show no higher-order dependency at all. The reason is that in diffusion processes such as the diffusion of information and the propagation of epidemics, according to the classic spreading models *(44)*, once a person *A* is infected, *A* will start to broadcast the information (or spread the disease) to all of its neighbors $\mathcal{N}(A)$, irrespective of who infected *A*. Due to this Markovian nature of diffusion processes, all diffusion data only show first-order dependencies and HON is identical to the first-order network. This also agrees with a previous finding that assuming second-order dependency has "marginal consequences for disease spreading" *(12)*.

**Improved accuracy on random walking**

Since random walking is a commonly used method to simulate movements on networks and is the foundation of many network analysis tools such as PageRank for ranking, MapEquation and walktrap for clustering, various link prediction algorithms and so on, it is crucial that a naïve random walker (only aware of the current node and its out-edges) can simulate the movements in the network accurately. If different network representations are built for the same sequential data set (consisting of trajectories), how will the network structure affect the movements of random walkers? Do the random walkers produce trajectories more similar to the real ones when running on HON?

We take the global shipping data to explain the experimental procedures (the clickstream data have similar results). As illustrated in Fig. 2A, for every trajectory of a ship, the last three locations are held for testing, and the others are used to construct the network. A first-order network (Fig. 2B), a fixed second-order network *(12)*, and a HON (Fig. 2C) are constructed from the same data set, respectively. Given one of the networks, for every ship, a random walker simulates the ship's movements on the network: It starts from the last location in the corresponding training trajectory, and walks three steps. Then the generated trajectories are compared with the ground truth in the testing set: a higher fraction of correct predictions means the random walkers can simulate the ship's movements better on the corresponding network. Random walking simulations in each network are repeated 1,000 times and the mean accuracies are reported. By comparing the accuracies of random walking, our intention here is not to solve a next location prediction problem *(21)* or similar classification problems, but from a network perspective, we focus on improving the representative power of the network, as reflected by the accuracies of random walking simulations.

The comparison of results among the conventional first-order network, the fixed second-order network, and HONs with maximum order of two to five are shown in Fig. 2D. It is shown that random walkers running on the conventional first-order network have significantly lower accuracies compared with other networks. The reason is that the first-order network representation only accounts for pairwise connections and cannot capture higher-order dependencies in ships' movement patterns. For example, a large proportion of ships are going back and forth between ports (e.g., Port *a* and Port *b* in Fig. 2A), which is naturally a higher-order dependency pattern because each ship's next step is significantly affected by its previous steps. Such return patterns are captured by HON (Fig. 2C), but not guaranteed in a first-order network (Fig. 2B where ships going from Port *a* to Port *b* may not return to Port *a*). As shown in Table 1, the probability of a ship returning to the same port after two steps in a first-order network (10.7%) is substantially lower than that in HON (above 40%). From another perspective, in a first-order network, a random walker is given more choices every step and is more "uncertain" making movements. Such "uncertainty" can be measured by the entropy rate *(12, 19)*, defined as:

$$H(X_{t+1} | X_t) = \sum_{i,j} \pi(i) p(i \to j) \log p(i \to j) \qquad (3)$$

where $\pi(i)$ is the stationary distribution at node $i$ and $p(i \to j)$ is the transition probability from node $i$ to node $j$, defined in Equation 1. The entropy rate measures the number of bits needed to describe every step of random walking — the more bits needed, the higher the uncertainty. In Table 1, the first-order network has the highest entropy rate, indicating that every step of random walking is more uncertain due to the lack of knowledge of what the previous steps are, which leads to the low accuracy in the simulation of movements.

By assuming an order of two for the whole network, the accuracies on the fixed second-order network increase considerably as in Fig. 2D, because the network structure can help the random walker remember its last two steps. Meanwhile, the accuracies on HON with a maximum order of two are comparable and slightly better than the fixed second-order network, because HON is able to capture second-order dependencies while avoid the overfitting caused by splitting all first-order nodes into second-order ones. Increasing the maximum order of HON can further improve the accuracy and lower the entropy rate; particularly, ship movements in bigger loops need more steps of memory and can only be captured with higher-orders, as reflected in Table 1, where the probability of returning in three steps increases from 7.3% to 16.4% when increasing the maximum order from two to three in HON. By increasing the maximum order to five, HON can capture all dependencies below or equal to the fifth order, and the accuracy of simulating one step on HON doubles that of the conventional first-order network.

Furthermore, when simulating multiple steps, the advantage of using HON is even bigger. The reason is that in a first-order network, a random walker "forgets" where it came from after each step, and has a higher chance of disobeying higher-order movement patterns. This error is amplified quickly in a few steps — the accuracy of simulating three steps on the first-order network is almost zero. On the contrary, in HON the higher-order nodes and edges can help the random walker remember where it came from, and provide the corresponding probability distributions for the next step. As a consequence, the simulation of three steps on HON is one magnitude more accurate than on first-order network. This indicates that when multiple steps are simulated (which is usually seen in methods such as PageRank and MapEquation that need multiple iterations), using HON (instead of the conventional first-order network) can help random walkers simulate movements more accurately, thus the results of all random walking-based network analysis methods will be more reliable.

**Effects on clustering**

One important family of network analysis methods is clustering, which identifies groups of nodes that are tightly connected. A variety of clustering algorithms such as MapEquation *(38)* and walktrap *(39)* are based on random walking, following the intuition that random walkers are more likely to move within the same cluster rather than between different clusters. Since using HON instead of a first-order network alters the movement patterns of random walkers running upon, a compelling question becomes: how does HON affect the clustering results?

Consider an important real-world application of clustering: identifying regions wherein aquatic species invasions are likely to happen. Since the global shipping network is the dominant global vector for the unintentional translocation of non-native aquatic species *(45)* (species get translocated either during ballast water uptake/discharge, or by accumulating on the surfaces of ships *(11)*), identifying clusters of ports that are tightly coupled by frequent shipping can reveal ports that are likely to introduce non-native species to each other. The limitation of the existing approach *(10)* is that the clustering is based on a first-order network that only accounts for direct species flows, while in reality the species introduced to a port by a ship may also come from multiple previous ports at which the ship has stopped due to partial ballast water exchanges and hull fouling. These indirect species introduction pathways driven by ship movements are already captured by HON and can influence the clustering result. As represented by the HON example in Fig. 1C, following the most likely shipping route, species are more likely to be introduced to Los Angeles from Shanghai (via Singapore) rather than from Tokyo, so the clustering (driven by random walking) on HON prefers grouping Los Angeles with Shanghai rather than with Tokyo. In comparison, indirect species introduction pathways are ignored when performing clustering on a first-order network (Fig. 1B), thus underestimating the risk of invasions via indirect shipping connections.

By clustering on HON, the overlap of different clusters is naturally revealed, highlighting ports that may be invaded by species from multiple regions. Since there can be multiple nodes representing the same physical location in HON (e.g., *Singapore|Tokyo* and *Singapore|Shanghai* both represent Singapore), and the ship movements through these nodes can be different, these higher-order nodes can belong to different clusters, so that Singapore as an international port belongs to multiple clusters, as one would expect.

The clustering results (using MapEquation) on a first-order network and HON are compared in Fig. 3. For example, let us consider Malta, a European island country in the Mediterranean Sea. Malta has two ports: Valletta is a small port that mainly serves cruise ships in Mediterranean, and Malta Freeport on the contrary is one of the busiest ports in Europe (many international shipping routes have a stop there). The clustering on the first-order network cannot tell the difference between the two ports and assigns both to the same Southern Europe cluster. On the contrary, the clustering on HON effectively separates Valletta and Malta Freeport by showing that Malta Freeport belongs to three additional clusters than Valletta, implying long-range shipping connections and species exchanges with ports all over the world. In summary, on HON, 45% of ports belong to more than one cluster, among which Panama Canal belongs to six clusters, and 44 ports (1.7% of all) belong to as many as five clusters, including international ports such as New York, Shanghai, Hong Kong, Gibraltar, Hamburg, and so on, indicating challenges to the management of aquatic invasions, as well as opportunities for devising targeted management policies. These insights are gained by adopting HON as the network representation for the global shipping data, while the MapEquation algorithm is unmodified.

**Effects on ranking**

Another important family of network analysis methods is ranking. PageRank *(37)* is commonly used in assessing the importance of web pages by using random walkers (with random resets) to simulate users clicking through different pages, and pages with higher PageRank scores have higher chances of being visited. It has been shown that web users are not Markovian *(14)*, and PageRank on the conventional network representation fails to simulate real user traffic *(46)*. Because HON can help random walkers achieve higher accuracies in reproducing movement patterns, how can HON affect the PageRank scores, and why?

With the clickstream data, we can construct both a first-order network and a HON as the input for PageRank. In HON, the PageRank scores of multiple higher-order nodes representing the same web page are summed up as the final score for the page. As shown in Fig. 4, by using HON instead of the first-order network, 26% of the web pages show more than 10% of relative changes in ranking; more than 90% of the web pages lose PageRank scores, while the other pages show remarkable gains in scores. To have an idea of the changes, we list the web pages that gain or lose the most scores by using HON as the input to PageRank, as shown in Table S1. Interestingly, of the 15 web pages that gain the most scores from HON, 6 are weather forecasts and 4 are obituaries, as one would expect considering this data set is from websites of local newspapers and TVs. Of the 15 web pages that lose the most scores, 3 are the lists of news personnel under the "about" page, which a normal reader will rarely visit, but over-valued by ranking on the first-order network.

To further understand how the structural differences of HON and the first-order network lead to changes in PageRank scores, we choose web pages that show significant changes in ranking, and compare the corresponding subgraphs of the two network representations. A typical example is a pair of pages, *PHOTOS: January 17th snow - WDBJ7 / news* and *View/Upload your snow photos - WDBJ7 / news* — these two pages gain 131% and 231% PageRank scores respectively on HON. In the first-order network representation, as shown in Fig. 5A where edge widths indicate the transition probabilities between web pages, it appears that after viewing or uploading the snow photos, a user is very likely to go back to the WDBJ7 home page immediately. In reality, however, once a user views and uploads a photo, the user is likely to repeat this process to upload more photos while less likely to go back to the home page. This natural scenario is completely ignored in the first-order network, but captured by HON, indicated by the strong loop between the two higher-order nodes (Fig. 5B). The example also shows how the higher probability of returning after two (or more) steps on HON can affect the ranking results. Again, all these insights are gained by using HON instead of the conventional first-order network, without any change to the PageRank algorithm. Besides the ranking of web pages, HON may also influence many other applications of ranking such as citation ranking and key phrase extraction.

**Scalability of HON**

We further show the scalability of HON, derived from its compact representation. In previous research (where a fixed second order is assumed for the network), from Table 1 it is shown that the network is considerably larger than the conventional first-order network, and assuming a fixed order beyond the second order becomes impractical because "higher-order Markov models are more complex" *(12)* due to combinatorial explosion. A network that is too large is computationally expensive to perform further analysis upon. On the contrary, while HON with maximum order of two has comparable accuracies in terms of random walking movement simulation, it has less nodes and about half the number of edges compared with a fixed second-order network, because it uses the first order whenever possible and embeds second-order dependencies only when necessary. Even when increasing the maximum order to five, HON still has less edges than the fixed second-order network, while all the useful dependencies up to the fifth order are incorporated in the network, resulting in considerably higher accuracies on random walking simulations.

Another important advantage of HON over a fixed-order network is that network analysis algorithms can run faster on HON, due to HON's compact representation. In addition, HON is sparser than the fixed-order representation, and many network toolkits are optimized for sparse networks. Table 1 shows the running time of two typical network analysis tasks: ranking (with PageRank *(37)*) and clustering (with MapEquation *(38)*). Compared with the fixed second-order network, these tasks run almost two times faster on HON with a maximum order of two, and about the same speed on HON with a maximum order of five (which embeds more higher-order dependencies and is more accurate).

It is worth noting that the number of additional nodes/edges needed for HON (on top of a first-order network) is determined by the number of higher-order dependencies in the data; that additional size is neither affected by the size of the raw data, nor the density of the corresponding first-order network. For example, even if the first-order network representation of a data set is a complete graph with 1 million nodes, if 100 second-order dependencies exist in the data, HON needs only 100 additional auxiliary second-order nodes on top of the first-order network, rather than making the whole network the second order. Thus, the advantage of HON is being able to effectively represent higher-order dependencies, while being compact by trimming redundant higher-order connections.

## Discussion

We have shown that for sequential data with higher-order dependencies, the conventional first-order network fails to represent such dependency patterns in the network structure, and the fixed second-order dependency can become limiting. If the network representation is not truly representative of the original data, then it will invariably lead to unreliable conclusions or insights from network analyses. We develop a new process for extracting higher-order dependencies in the raw data, and for building a network (the Higher-Order Network (HON)) that can represent such higher-order dependencies. We demonstrate that our novel network representation is more accurate in representing the true movement patterns in data in comparison with the conventional first-order network or the fixed second-order network: for example, when using HON instead of a first-order network to represent the global shipping data, the accuracy is doubled when simulating a ship's next movement on the network, and is higher by one magnitude when simulating three steps, since the higher-order nodes and edges in HON can provide more detailed guidance for simulated movements. Besides improved accuracy, HON is more compact than fixed-order networks by embedding higher-order dependencies only when necessary, and thus network analysis algorithms run faster on HON.

Furthermore, we show that using HON instead of conventional network representations can influence the results of network analyses methods that are based on random walking. For example, on HON, the clustering of ports takes indirect ship-borne species introduction pathways into account, and naturally produces overlapping clusters that indicate multiple sources of species invasion for international ports; the ranking of web pages is corrected by incorporating the higher-order patterns of users' browsing behaviors such as uploading multiple photos. Our work has the potential to influence a wide range of applications, such as improving PageRank for the task of unsupervised key phrase selection in language processing *(47)*, as the proposed network representation is consistent with the input expected by various network analysis methods. Since nodes could be split into multiple ones in HON, it may require post-processing to aggregate the results for interpretation. In the current method, the choice of parameters may influence the structure of the resulting network, so we provide parameter discussions in Supplementary Material Note 3.

In future work, we look forward to (a) extending the applications of HON beyond the simulation of movements to more dynamic processes such as dynamic network anomaly detection *(48)*, and (b) improving the algorithm by reducing the parameters needed.

## Materials and Methods

The construction of the Higher-Order Network (HON) consists of two steps: *rule extraction* identifies higher-order dependencies that have sufficient support and can significantly alter a random walker's probability distribution of choosing the next step; then *network wiring* adds these rules describing variable orders of dependencies into the conventional first-order network by adding higher-order nodes and rewiring edges. The data structure of the resulting network is consistent with the conventional network representation, so existing network analysis methods can be applied directly without being modified. We use global shipping traffic data as a working example to demonstrate the construction of HON, but it is generalizable to any sequential data.

**Rule extraction**

The challenge of rule extraction is to identify the appropriate orders of dependencies in data; when building the first-order network, this step is often ignored by simply counting pairwise connections in the data. We define a *path* as the movement from source node $A$ to target node $B$, though with nodes that differ from those in a conventional network: a node here can represent a sequence of entities, no longer necessarily a single entity. Then among those paths, given a source node $A$ containing a sequence of entities $[a_1, a_2, …, a_k]$, if including an additional entity $a_0$ at the beginning of $A$ can significantly alter the normalized counts of movements (as probability distribution) to target nodes set $\{B\}$, it means $\{B\}$ has a higher-order dependency on $A_{\text{ext}} = [a_0, a_1, a_2, …, a_k]$, and paths containing higher-order dependencies like $A_{\text{ext}} \to B$ are defined as *rules*. Then a rule like Freq($[a_0, a_1] \to a_2$) = 50 can map to an edge in the network in the form of $a_1|a_0 \to a_2$ with edge weight 50. What are the expectations for the rule extraction process?

First, rules should represent dependencies that are significant. As in Figure S1③, if the probability distribution of a ship's next step from Singapore is significantly affected by knowing the ship came from Shanghai to Singapore, there is at least second-order dependency here. On the contrary, if the probability distribution of going to the next port is the same no matter how the ship reached Singapore, there is no evidence for second-order dependency (but third or higher-order dependencies may still exist, such as $g|f,d$ in Fig. S4C, and can be checked similarly).

Second, rules should have sufficient support. Only when some pattern happens sufficiently many times can it be considered as a "rule" rather than some random event. Although this requirement of minimum support is not compulsory, not specifying a minimum support will result in a larger and more detailed network representation, and more infrequent routes are falsely considered as patterns, ultimately lowering the accuracy of the representation (see the discussions of parameters in Supplementary Material Note 3).

Third, rules should be able to represent variable orders of dependencies. In real-world data such as the global shipping data, different paths can have different orders of dependencies, for example in Fig. 1D the next step from Singapore is dependent on Tianjin through the fourth-order path *Tianjin → Busan → Tokyo → Singapore*, as well as on Shanghai through the second-order path *Shanghai → Singapore*. When variable orders can co-exist in the same data set, the rule extraction algorithm should not assign a fixed order to the data, but should be able to yield rules representing variable orders of dependencies.

Following the aforementioned three objectives of rule extraction, it is natural to *grow* rules incrementally: start with a first-order path, try to increase the order by including one more previous step, and check if the probability distribution for the next step changes significantly (Fig. S1③). If the change is significant, the higher order is assumed, otherwise keep the old assumption of order. This rule growing process is iterated recursively until (a) the minimum support requirement is not met, or (b) the maximum order is exceeded. The detailed algorithm is given in the Supplementary Material Note 2.

**Network wiring**

The remaining task is to convert the rules obtained from the last step into a graph representation. It is trivial for building conventional first-order networks because every rule is first-order and can directly map to an edge connecting two entities, but such direct conversion will not work when rules representing variable orders of dependencies co-exist. The reason is that during rule extraction, only the last entity of every path is taken as the target node, so that every edge points to a first-order node, which means higher-order nodes will not have in-edges. Rewiring is needed to ensure that higher-order nodes will have incoming edges, while preserving the sum of edge weights in the network. The detailed steps are illustrated as follows:

1. Converting all first-order rules into edges. This step is exactly the same as constructing a first-order network, where every first-order rule (a path from one entity to another) corresponds to a weighted edge. As illustrated in Fig. S2A, *Shanghai → Singapore* is added to the network.

2. Converting higher-order rules. In this step, higher-order rules are converted to higher-order edges pointing out from higher-order nodes (the nodes are created if they do not already exist in the network). Fig. S2B shows the conversion of rules *Singapore|Shanghai → Los Angeles* and *Singapore|Shanghai → Seattle*, where the second-order node *Singapore|Shanghai* is created and two edges pointing out from the node are added.

3. Rewiring in-edges for higher-order nodes. This step preserves the sum of edge weights while solving the problem that higher-order nodes have no incoming edges, by pointing existing edges to higher-order nodes. When adding the second-order node *Singapore|Shanghai*, a lower order rule and the corresponding edge *Shanghai → Singapore* are guaranteed to exist, because during rule extraction when a rule is added, all preceding steps of the path are also added, as in ADDTORULES in Algorithm 1. As shown in Fig. S2C, the edge from *Shanghai* to *Singapore* is redirected to *Singapore|Shanghai*. Converting higher-order rules (Step 2) and rewiring (Step 3) are repeated for all rules of first order, then second order, and likewise up to the maximum order, in order to guarantee that edges can connect to nodes with the highest possible orders. This step also implies that any two nodes that represent the same physical location will not have incoming edges from the same node.

4. Rewiring edges built from *Valid* rules. This step after representing all rules as edges in HON is necessary due to that the rule extraction step takes only the last entity of paths as targets, such that edges built from *Valid* rules in Algorithm 1 always point to first-order nodes. In Fig. S2D, the node *Singapore|Shanghai* was pointing to a first-order node *Seattle*. However, if a node of higher order *Seattle|Singapore* already exists in the network, the edge *Singapore|Shanghai → Seattle* should point to *Seattle|Singapore*, otherwise the information about previous steps is lost. To preserve as

much information as possible, the edges built from *Valid* rules should point to nodes with the highest possible orders.

Following the above process, the algorithm for network wiring is given in Algorithm 2 in Supplementary Material Note 2, along with more detailed explanations. Given a set of parameters, the result of HON is unique, so there is no optimization or greedy methods for the algorithm. Note that we have made the entire source code available as well (at https://github.com/xyjprc/hon).

**Comparison with related methods**

Although Variable-order Markov (VOM) models can be used on sequential data to learn a VOM tree *(22–24)* for predictions *(49)*, our goal is to build a more accurate network representation that captures higher-order dependencies in the data. While these two objectives are related, there are several key differences: (a) a VOM tree contains probabilities that are unnecessary (e.g. nodes that are not leaves) for representing higher-order dependencies in a network; (b) additional conditional probabilities are needed to connect nodes with different orders in HON, which are not guaranteed to exist in a pruned VOM tree; (c) VOM usually contains lots of unnecessary edges due to the "smoothing" process for the unobserved data, which is not desired for a network representation. Therefore, our work is not simply contained in a VOM implementation. Supplementary Material Note 4 elaborates the differences and provides an empirical comparison between the HON and VOM.

Although a fixed $k^{th}$-order Markov model can be directly converted to a first-order model *(12)*, the state space $S^k$ grows exponentially with the order. There has been plenty of research on Markov order estimation to determine the order $k$ such as *(26, 27)* using different information criteria, *(14)* using cross validation, and *(28)* using surrogate data, but these approaches produce a single global order for the model rather than variable orders, and no discussion was given to network representation. Other Markov-related works such as hidden Markov model *(25)*, frequent pattern mining *(20)*, and next location prediction *(21)* focus more on the stochastic process, rather than the network representation problem. For example, the hidden states in HMM do not represent clear dependency relationships like the higher-order nodes do in HON, and we are not learning a hidden layer that have "emission probabilities" to observations. From the network perspective, while there have been efforts to incorporate higher-order network structures for clustering *(34, 35)*, ranking *(36)*, and so on, these methods are modifications of existing algorithms and are application-specific; instead, we embed higher-order dependencies into the network structure, so that the wide range of existing network analysis tools can be applied without modification.

# Supplementary Material

Data sets

Algorithms

Parameter discussion

Empirical comparison with the Variable-order Markov (VOM) model

**Acknowledgements**: We thank David Lodge, Yuxiao Dong, and Reid Johnson for their valuable comments.

**Funding**: This work is based on research supported by the ND Office of Research via Environmental Change Initiative (ECI), the National Science Foundation (NSF) Award EF-1427157, IIS-1447795, and BCS-1229450, and the Army Research Laboratory and was accomplished under Cooperative Agreement Number W911NF-09-2-0053.

**Author contributions**: JX, TLW and NVC collectively conceived the research. JX, TLW, and NVC designed the analyses. JX conducted the experiments. JX, TLW and NVC conducted the analyses. JX, TLW and NVC wrote the manuscript.

**Competing interests**: The authors declare that they have no competing financial interests.

**Data and materials availability**: The entire source code and the synthetic data are available online at https://github.com/xyjprc/hon.

## Figures and Tables

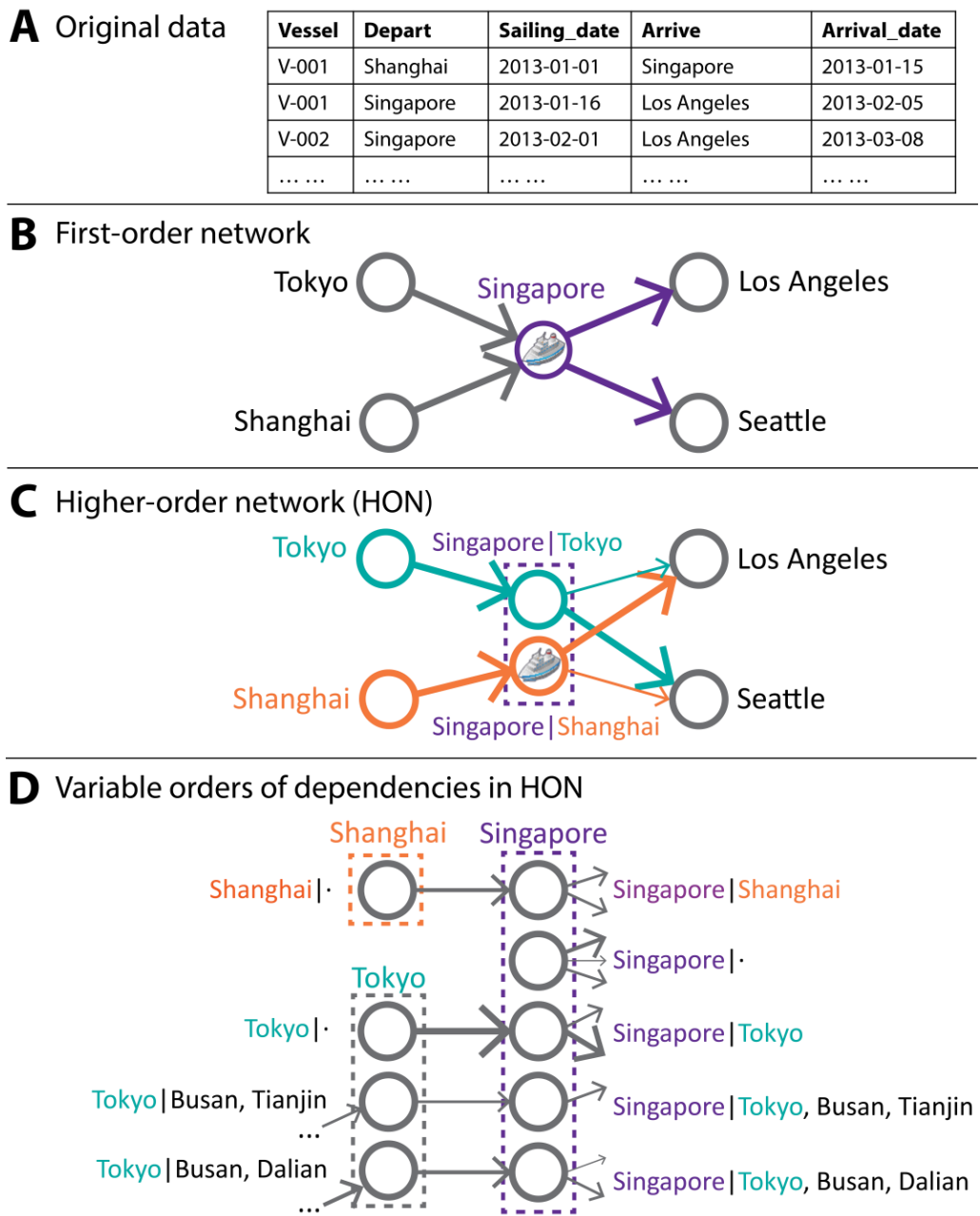

**Fig. 1. Necessity of representing dependencies in networks.** (**A**) A global shipping data set, containing ship movements as sequential data. (**B**) A first-order network built by taking the number of trips between port pairs as edge weights. A ship currently at Singapore has similar probabilities of going to Los Angeles and Seattle, no matter where the ship came to Singapore. (**C**) By breaking down the node Singapore, the ship's next step from Singapore can depend on where the ship came to Singapore, thus more accurately simulate movement patterns in the original data. (**D**) Variable orders of dependencies represented in HON. First-order to fourth-order dependencies are shown here, and can easily extend to higher orders. Coming from different paths to Singapore, a ship will choose the next step differently.

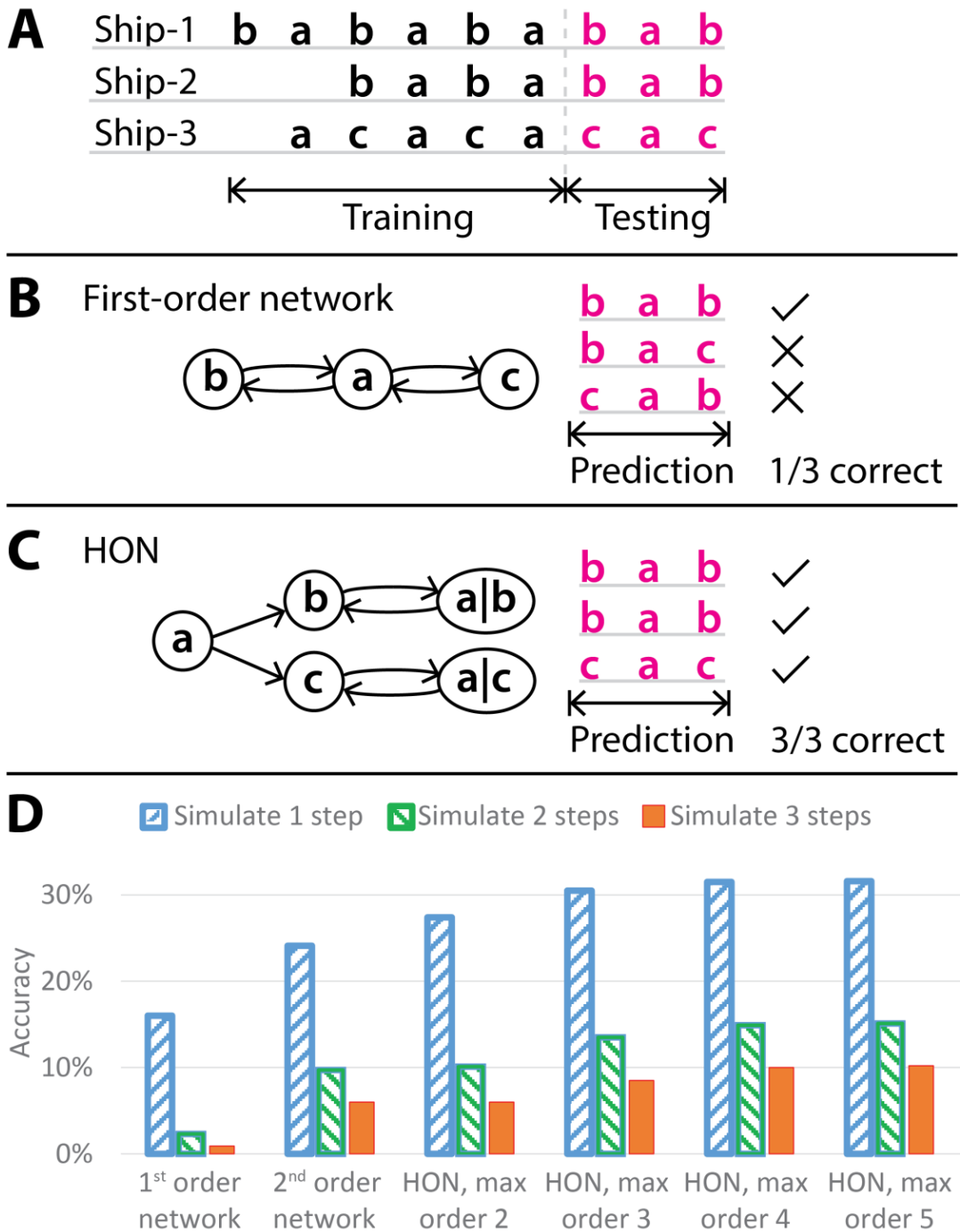

**Fig. 2. Comparison of random walking accuracies.** (**A**) For the global shipping data comprised of ships' trajectories, hold the last three steps of each trajectory for testing and use the rest to build the network. (**B** and **C**) Given a generated shipping network, every ship is simulated by a random walker, which walks three steps from the last location in the corresponding training trajectory. The generated trajectories are compared with the ground truth, and the fraction of correct predictions is the random walking accuracy. (**D**) By using HON instead of the first-order network, the accuracy is doubled when simulating the next step, and improved by one magnitude when simulating the next three steps. Note that error bars are too small to be seen (standard deviations on HONs are 0.11% ±0.02%).

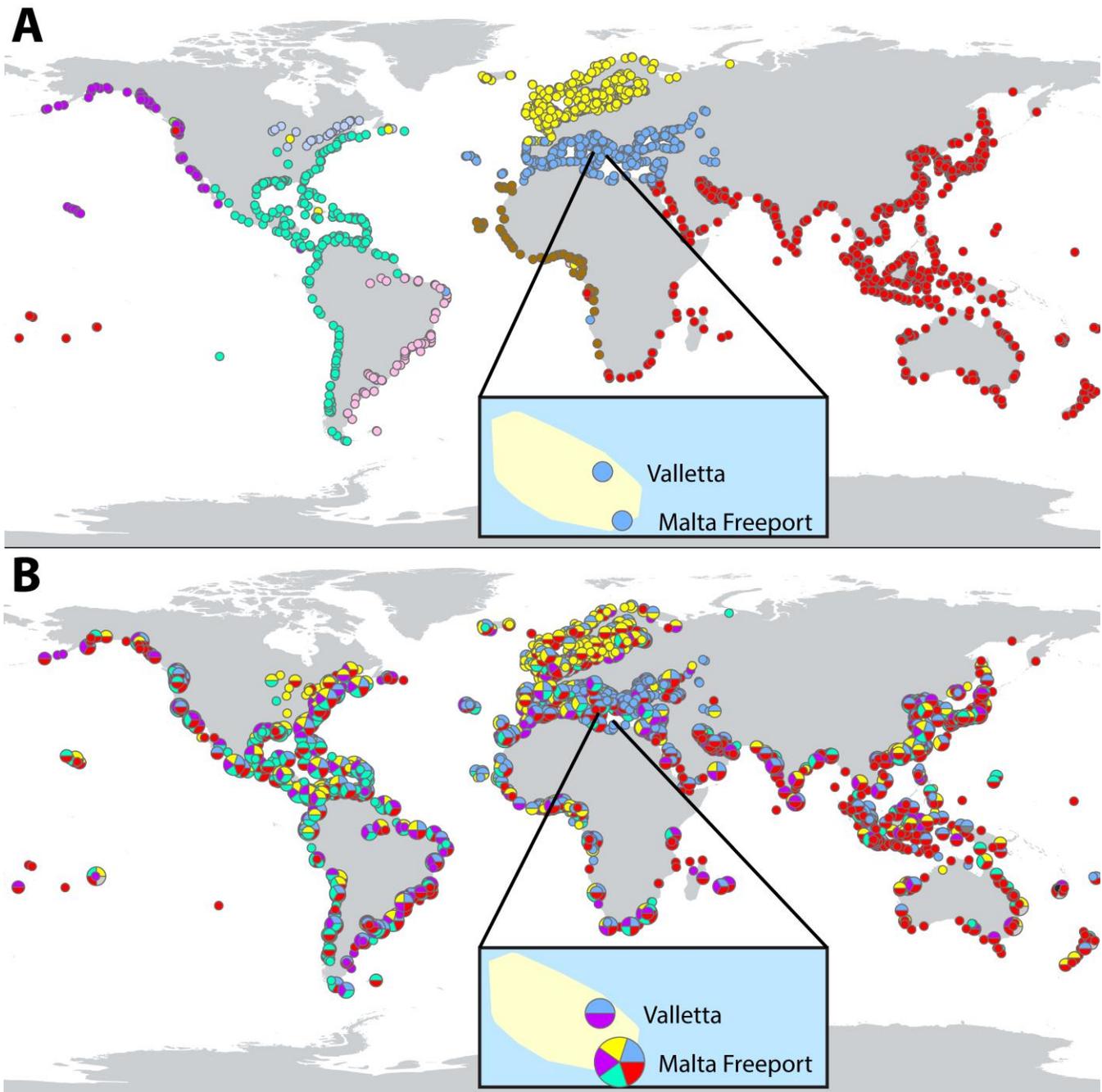

**Fig. 3. Clustering of ports on different network representations of the global shipping data.** Ports tightly coupled by frequent shipping in a cluster are likely to introduce non-native species to each other. MapEquation *(38)* is used for clustering, and different colors represent different clusters. (**A**) Clustering on the first-order network. Although Valletta and Malta Freeport are local and international ports respectively, the clustering result does not distinguish the two. (**B**) Clustering on HON. The overlapping clusters indicate how international ports (such as Malta Freeport) may suffer from species invasions from multiple sources.

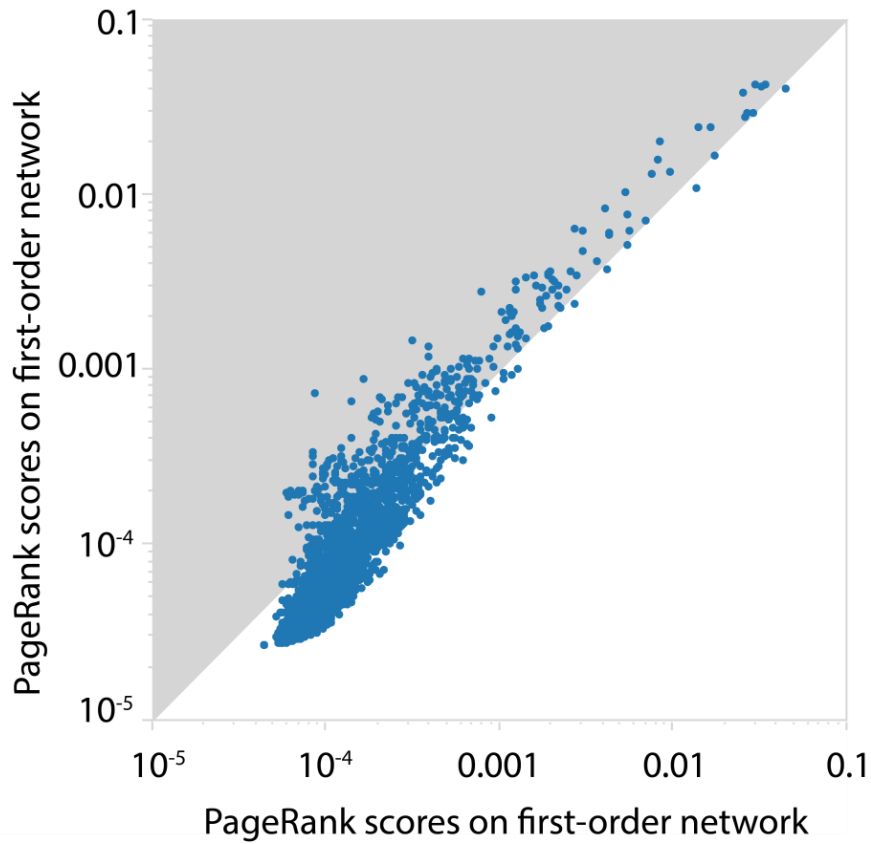

**Fig. 4. Change of web page rankings by using HON instead of first-order network.** PageRank *(37)* is used for ranking. 26% of the pages show more than 10% of relative changes in ranking. More than 90% of the web pages lose PageRank scores, while the other pages show remarkable gain in scores. Note that log-log scale is used in the figure, so a deviation from the diagonal indicates a significant change of the PageRank score.

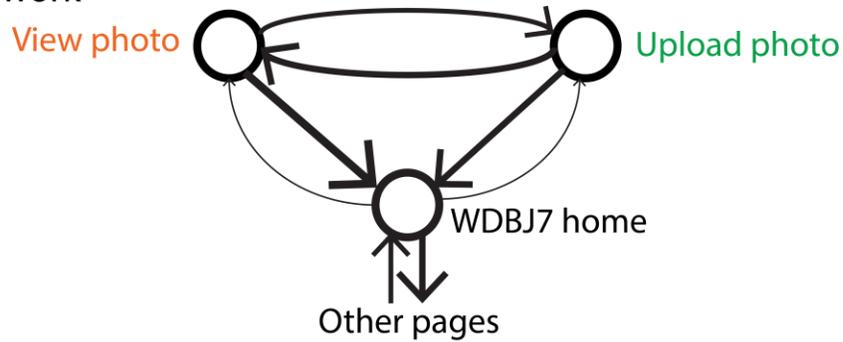
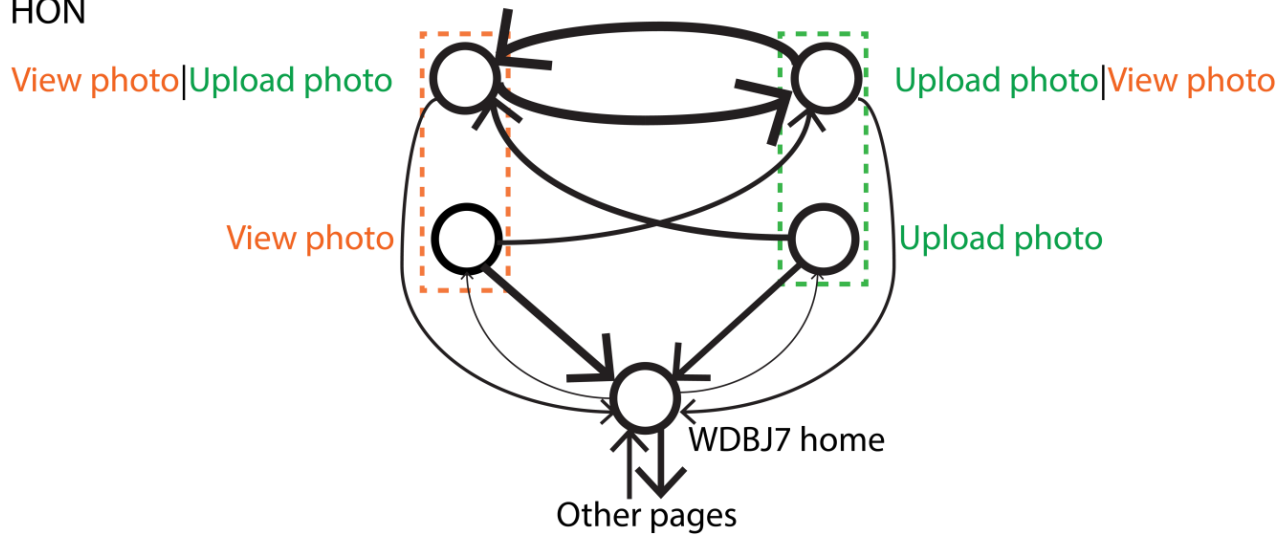

**Fig. 5. Comparison of different network representations for the same clickstream data.** Edge widths indicate the transition probabilities. (**A**) The first-order network representation, indicating a user is likely to go back to the homepage after viewing or uploading snow photos. (**B**) The HON representation, which not only preserves the information in the first-order network, but also uses higher-order nodes and edges to represent an additional scenario: once a user views and uploads a photo, the user is likely to repeat this process to upload more photos and is less likely to go back to the home page. Consequently, these photo viewing and uploading pages will receive higher PageRank scores *(37)* because the implicit random walkers of PageRank are more likely to be trapped in the loop of the higher-order nodes.

**Table. 1. Comparing different network representations of the same global shipping data.**

| Network representation | Number of edges | Number of nodes | Network density | Prob. of returning after two steps | Prob. of returning after three steps | Entropy rate (bits) | Clustering time (mins) | Ranking time (s) |
|---|---|---|---|---|---|---|---|---|
| Conventional first-order | 31,028 | 2,675 | $4.3 \times 10^{-3}$ | 10.7% | 1.5% | 3.44 | 4 | 1.3 |
| Fixed second-order | 116,611 | 19,182 | $3.2 \times 10^{-4}$ | 42.8% | 8.0% | 1.45 | 73 | 7.7 |
| HON, max order two | 64,914 | 17,235 | $2.2 \times 10^{-4}$ | 41.7% | 7.3% | 1.46 | 45 | 4.8 |
| HON, max order three | 78,415 | 26,577 | $1.1 \times 10^{-4}$ | 45.9% | 16.4% | 0.90 | 63 | 6.2 |
| HON, max order four | 83,480 | 30,631 | $8.9 \times 10^{-5}$ | 48.9% | 18.5% | 0.68 | 67 | 7.0 |
| HON, max order five | 85,025 | 31,854 | $8.4 \times 10^{-5}$ | 49.3% | 19.2% | 0.63 | 68 | 7.6 |

# Supplementary Material

## 1 Data sets

**Global shipping data.** This data made available by Lloyd's Maritime Intelligence Unit (LMIU) contains ship movement information such as `vessel_id`, `port_id`, `sail_date` and `arrival_date`. Our experiments are based on a recent LMIU data set that spans one year from May 1st, 2012 to April 30th, 2013, totaling 3,415,577 individual voyages corresponding to 65,591 ships that move among 4,108 ports and regions globally. A minimum support of 10 is used to filter out noise in the data.

**Clickstream data.** This data made available by a media company contains logs of users clicking through web pages that belong to 50 news web sites owned by the company. Fields of interest include `user_ip`, `pagename` and `time`. Our experiments are based on the clickstream records that span two months from December 4th, 2012 to February 3rd, 2013, totaling 3,047,697 page views made by 179,178 distinct IP addresses on 45,257 web pages. A minimum support of 5 is used to filter out noise in the data. Clickstreams that are likely to be created by crawlers (abnormally long clickstreams / clickstreams that frequently hit the error page) are omitted.

**Retweet data.** This data *(50)* records retweet history on Weibo (a Chinese microblogging website), with information about who retweets whose messages at what time. The data was crawled in 2012 and there are 23,755,810 retweets recorded, involving 1,776,950 users.

**Synthetic data**. We created a trajectory data set (data and code are available at https://github.com/xyjprc/hon) with known higher-order dependencies to verify the effectiveness of the rule extraction algorithm. In the context of shipping, we connect 100 ports as a 10×10 grid, then generate trajectories of 100,000 ships moving among these ports. Each ship moves 100 steps, yielding 10,000,000 movements in total. Normally each ship has equal probabilities of going up/down/left/right on the grid in each step (with wrapping, e.g., going down at the bottom row will end up in the top row); we use additional higher-order rules to control the generation of ship movements. For example, a second-order rule can be defined as whenever a ship comes from Shanghai to Singapore, instead of randomly picking a neighboring port of Singapore for the next step, the ship has 70% chance of going to Los Angeles and 30% chance of going to Seattle. We predefine 10 second-order rules like this, and similarly 10 third-order rules, 10 fourth-order rules, and no other higher-order rules, so that movements that have variable orders of dependencies are generated. To test the rule extraction algorithm, we set the maximum order as five to see if the algorithm will incorrectly extract false rules beyond the fourth order which we did not define; we set minimum support as five for patterns to be considered as rules.

## 2  Algorithms

**Rule extraction.** Algorithm 1 gives the pseudocode of rule extraction. The procedure consists of three major steps: BUILDOBSERVATIONS counts the frequencies of all subsequences from the second order to *MaxOrder* for every trajectory in *T* (Fig. S1 ①); BUILDDISTRIBUTIONS first builds all paths by removing subsequences that appear less than *MinSupport* times, then estimates probability distributions of movements at every source node by normalizing the observed frequencies (Fig. S1 ②). GENERATEALLRULES starts from the first order and tries to increase the order recursively, by including an additional entity at the beginning of the entity sequence of the source node and testing if the probability distribution for the next step changes significantly (Fig. S1 ③).

The comparison between probability distributions is performed by the recursive function EXTENDRULE. *Curr* is the current source node ($X_t$ = *Singapore* in Fig. S1), which is to be extended into a higher-order source node *ExtSource* by including the previous step ($X_t$ = *Singapore*, $X_{t-1}$ = *Shanghai* in Fig. S1) (line 40). *Valid* is the last known source node from which a random walker has significantly different probability distributions towards the next step, i.e., the last assumed order for the path starting from node *Valid* is length(*Valid*). If at the extended source node, a random walker has a significantly different probability distribution of the next movement compared with that at node *Valid*, the extended source node will be marked as the new *Valid* for the recursive growing of rule (line 47), otherwise the old *Valid* is kept (line 49). The paths with *Valid* as the source node have correct orders of dependencies, and will be added to the rules set *R* whenever EXTENDRULE exceeds *MaxOrder* (line 35) or the source node cannot be extended (line 41, true when no higher-order source node with the same last steps exists). When a higher-order rule (a path from *Source*) is added, all paths of the preceding steps of *Source* are added (line 54–55) to ensure the network wiring step can connect nodes with variable orders. For example, when paths from the source node *Singapore|Shanghai* are added to *R*, the preceding step *Shanghai* → *Singapore* should also be added to *R*.

Specifically, to determine whether extending the source node from *Valid* to *ExtSource* significantly changes the probability distribution for the next step (line 46), we compute the Kullback–Leibler divergence *(51)* between the two distributions, since it is a widely-used and standard way of comparison probability distributions *(22–24)*. We consider the change is significant if the divergence satisfies

$Divergence_{KL}(P_{Valid}, P_{ExtSource}) > \frac{Order_{ExtSource}}{\log_2(Support_{ExtSource})}$, based on the following intuition, inspired by Ben-Gal 2005 *(24)* and Bühlmann 1999 *(22)*: for the two extended nodes showing the same divergence with regard to the original source node, (1) we are more inclined to prune the node with higher orders (the one with more previous steps embedded); and (2) we are less inclined to prune the node with higher support (the one with more trajectories going through). Instead of applying a universal threshold for all nodes that may have varying orders and supports, the threshold we adopt is self-adjustable for different nodes.

It is worth noting that Algorithm 1 also applies to data types other than trajectories such as diffusion data, which needs only one change of the EXTRACTSUBSEQUENCES function such that it takes only the newest entity subsequence. In addition, although higher-order dependencies exist in many types of data, it is the type of data (vessel trajectory data / gene sequence data / language data / diffusion data …) that determines whether there are higher-order dependencies and how high the orders can be. Our proposed algorithm is backwards compatible with data that have no higher-order dependencies (such as the diffusion data): all rules extracted are first-order, thus the output will be a first-order network.

**Network wiring.** Algorithm 2 gives the pseudocode for converting rules extracted from data into a graph representation. In line 3, sorting rules by the length of key *Source* is equivalent to sorting rules by ascending orders. This ensures that the for loop in line 4–7 converts all lower order rules before processing higher-order rules. For every order, line 5 converts rules to edges, and line 7 REWIRE(*r*) attempts rewiring if it is not the first order. Fig. S2C illustrates the case *r = Singapore|Shanghai → Los Angeles* in line 11, indicating *PrevSource* is *Shanghai* and *PrevTarget* is *Singapore|Shanghai*. The edge *Shanghai → Singapore|Shanghai* is not found in the network, so in line 15 and 16 *Shanghai → Singapore|Shanghai* replaces *Shanghai → Singapore* using the same edge weight.

After converting all rules in REWIRETAILS, edges created from *Valid* rules are rewired to target nodes that have the highest possible orders. Fig. S2D illustrates the following case: in line 21 *r = Singapore|Shanghai → Seattle*, so in line 22 *NewTarget* is assigned as *Seattle|Singapore,Shanghai*; assume *Seattle|Singapore,Shanghai* is not found in all sources of *R*, so the lower order node *Seattle|Singapore* is searched next; assume *Seattle|Singapore* already exists in the graph, so *Singapore|Shanghai → Seattle|Singapore* replaces *Singapore|Shanghai → Seattle*.

# 3   Parameter discussion

**Minimum support.** As mentioned in the methods section, only when a subsequence occurs sufficiently many times (not below minimum support) can it be distinguished from noise and construed as a (non-trivial) path. While a minimum support is not compulsory, setting an appropriate minimum support can significantly reduce the network size and improve the accuracy of representation. As shown in Fig. S3A, by increasing the minimum support from 1 to 10 (with a fixed maximum order of 5), the size of the network shrinks by 20 times while the accuracy of random walking simulation increases by 0.54%. By increasing the minimum support from 1 to 100, the accuracy first increases then decreases. The reason is that with low minimum support, some unusual subsequences that are noise are counted as paths; on the other hand, a high minimum support leaves out some true patterns that happen less frequently. The optimal minimum support (that can increase the accuracy of representation and greatly reduce the size of the network) may not be the same for different types of data, but can be found by parameter sweeping.

**Maximum order.** With a higher maximum order, the rule extraction algorithm can capture dependencies of higher orders, leading to higher accuracies of random walking simulations. As shown in Fig. S3B, when increasing the maximum order from 1 to 5 (with a fixed minimum support of 10), the accuracy of random walking simulation keeps increasing but converges at the maximum order of 5, and the same trend applies for the size of the network. The reason is that the majority of dependencies have lower orders while fewer dependencies have higher orders, which again justifies our approach of not assigning a fixed high order for the whole network. On the other hand, setting a high maximum order does not significantly increase the running time of building HON, because in the rule extraction algorithm, most subsequences of longer lengths do not satisfy the minimum support requirement and are not considered in the following steps. In brief, when building HON using the aforementioned algorithm, an order that is sufficiently high can be assumed as the maximum order (a maximum order of five is sufficient for most applications).

# 4   Empirical comparison with the Variable-order Markov (VOM) model

In Materials and Methods we mentioned the differences between HON and the Variable-order Markov (VOM) model *(22–24)*. In this section we first illustrate these differences with an example, then provide an empirical comparison between HON and VOM. Note that because the "smoothing" process is not a compulsory step of VOM, we do not apply it in the following comparison, although "smoothing" is undesirable for network representation.

**Example**. In the context of global shipping, suppose ports *a*, *b*, *c*, …, *h*, *i* are connected as shown in Fig. S4A. Port *f* and *g* are at the two ends of a canal. We assume that all ships coming from *d* through the canal will go to *h*, and all ships coming from *e* through the canal will go to *i*. A possible set of ship trajectories are listed in Fig. S4B. Based on these trajectories, we can count the frequencies of subsequences, and compute the probability distributions of next steps given the previous ports visited (HON and VOM yield identical results). The subsequences of variable orders can naturally form a tree as shown in Fig. S4C, where source nodes are in circles and target nodes (and the corresponding edge weights) are in the boxes below.

HON and VOM have different mechanisms of deciding which nodes to retain in the tree. In Fig. S4C, the nodes kept by HON are denoted by red stars and nodes kept by VOM are denoted by purple triangles, showing a mismatch of the results. For HON, although *g*|*f* does not show second order dependency (having the same probability distribution with *g*), *g*|*f,d* shows third order dependency (having significantly different probability distribution compared with g), so *g*|*f,d* is retained by HON. According to ADDTORULES of the "rule extraction" step (Algorithm 1), all preceding nodes are retained, including *f*|*d* and *d*, such that the "network wiring" step already has exactly the nodes needed: there would be a path of *d* → *f*|*d* → *g*|*f,d*, as shown in Fig. S4D. On the contrary, in the VOM construction process, after determining that *g*|*f,d* is a higher-order node to be kept, VOM keeps *g*|*f*, and prunes *f*|*d*, despite that (1) *f*|*d* is necessary for building the link to *g*|*f,d* when constructing a network, and (2) *g*|*f* is not necessary for building HON as it has the same probability distribution with g.

The eventual wiring of HON is shown in Fig. S4D. Compared with the true connection in Fig. S4A, HON not only keeps the first order links, but also adds higher-order nodes and edges for the two ports *f* and *g* in the canal, successfully capturing the pattern that "all ships coming from *d* through the canal will go to *h*, and all ships coming from *e* through the canal will go to *i*".

An additional difference between HON and VOM is how they determine the orders of rules. HON assumes the first order initially and compares with higher orders, while VOM "prunes" rules recursively from higher orders to lower orders, which as illustrated in Fig. S4E, may prune higher-order nodes despite they have very different distributions than first-order nodes (e.g., *z*|*y,x,w* compared with *z*), thus underestimating the orders of dependencies.

In brief, we have shown that VOM cannot be used directly to construct HON, given that VOM (1) retains unnecessary nodes for constructing HON, (2) prunes necessary nodes, and (3) has a pruning mechanism that may leave out certain higher-order dependencies.

**Numerical comparison**. To show the differences of HON and VOM quantitatively, we apply both HON and VOM to the same global shipping data set, assume the same filtering for preprocessing

(*MaxOrder* = 5 and *MinSupport* = 10), use the same distance measure (KL divergence), and for fair comparison, we use the same threshold $\frac{Order_{ExtSource}}{\log_2(Support_{ExtSource})}$ for judging whether two distributions are significantly different. Table S2 gives the comparison of the number of rules extracted from both algorithms.

We can observe that the rules extracted by HON and VOM show considerable differences except for the first order, even though these two algorithms are given the same parameters. The different mechanisms of deciding which nodes to keep lead to the differences in the extracted rules. This further supports our claim that the rules extracted by VOM cannot be readily used for building HON, while the "rule extraction" process of HON has already prepared exactly the rules needed and only need to rewire some links.

# Supplementary figures, tables, and algorithms

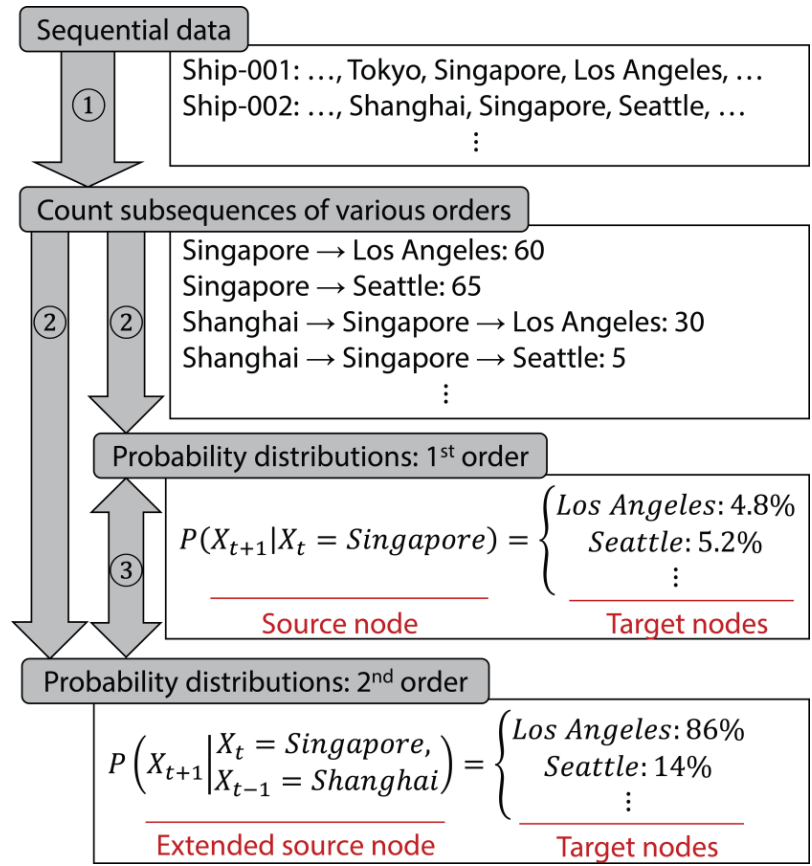

**Fig. S1. Rule extraction example for the global shipping data.** Step 1: count the occurrences of subsequences from the first order to the maximum order, and keep those that meet the minimum support requirement. Step 2: given the source node representing a sequence of entities as the previous step(s), compute probability distributions for the next step. Step 3: given the original source node and an extended source node (extended by including an additional entity at the beginning of the entity sequence), compare the probability distributions of the next step. For example, when the current location is Singapore, knowing that a ship comes from Shanghai to Singapore (second order) significantly changes the probability distribution for the next step compared with not knowing where the ship came from (first order). So the second-order dependency is assumed here; then the probability distribution is compared with that of the third order, and so on, until the minimum support is not met or the maximum order is exceeded.

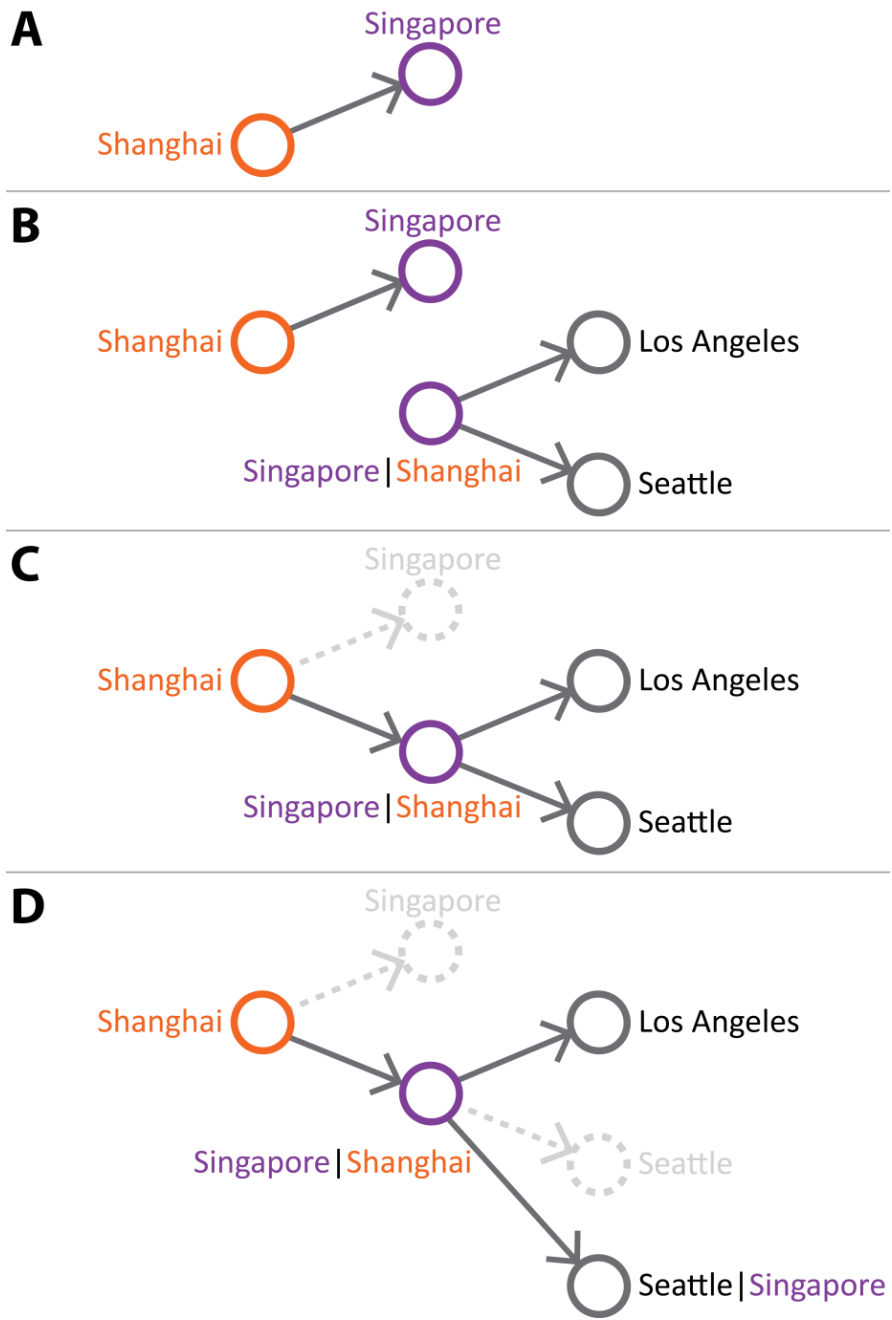

**Fig. S2. Network wiring example for the global shipping data.** Figure shows how the dependency rules are represented as HON. (**A**) convert all first-order rules into edges; (**B**) convert higher-order rules, and add higher-order nodes when necessary, (**C**) rewire edges so that they point to newly added higher-order nodes (the edge weights are preserved); (**D**) rewire edges built from Valid rules so that they point to nodes with the highest possible order.

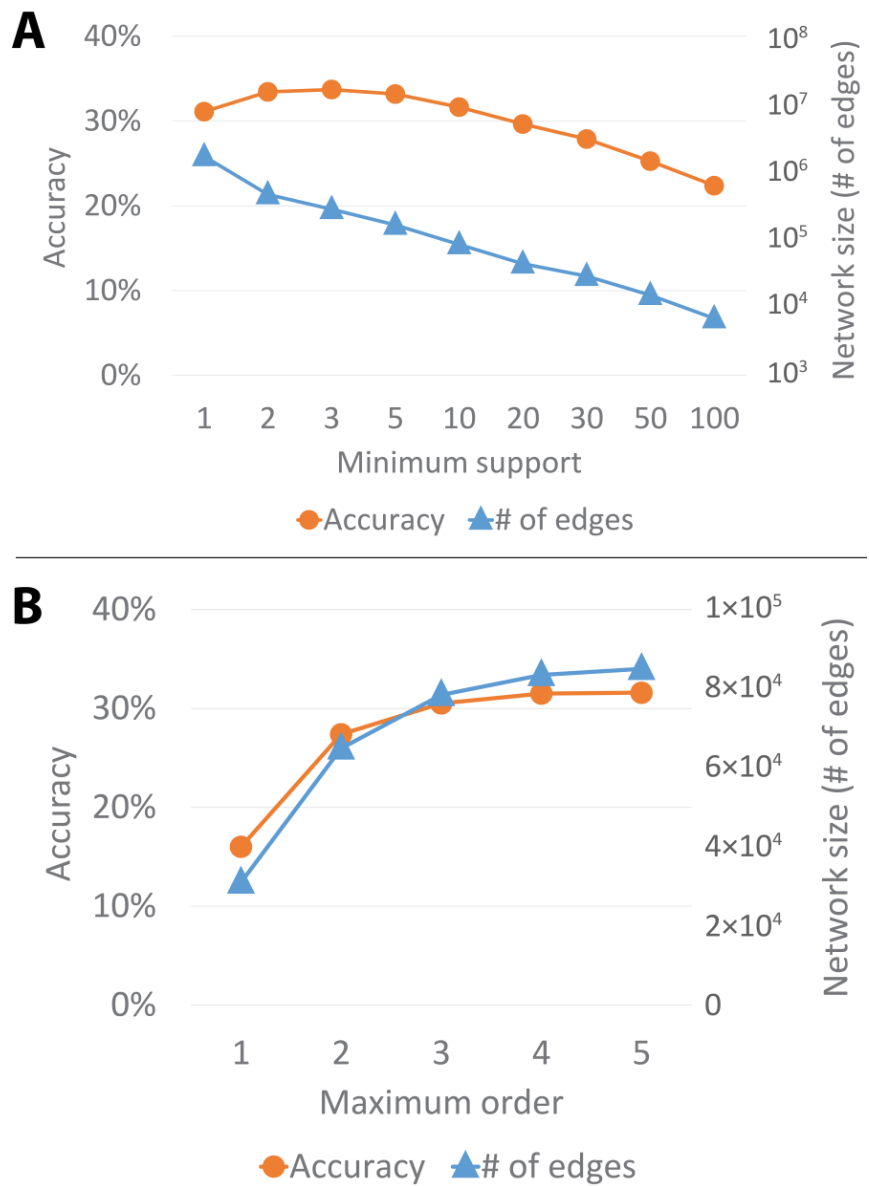

**Fig. S3. Parameter sensitivity of HON in terms of the accuracy and network size.** The global shipping data is illustrated, and the accuracy is the percentage of correct predictions when using a random walker to predict the next step. (**A**) An appropriate minimum support can significantly reduce the network size and improve the accuracy of representation; (**B**) when increasing the maximum order, the accuracy of random walking simulation keeps improving but converges near the maximum order of 5.

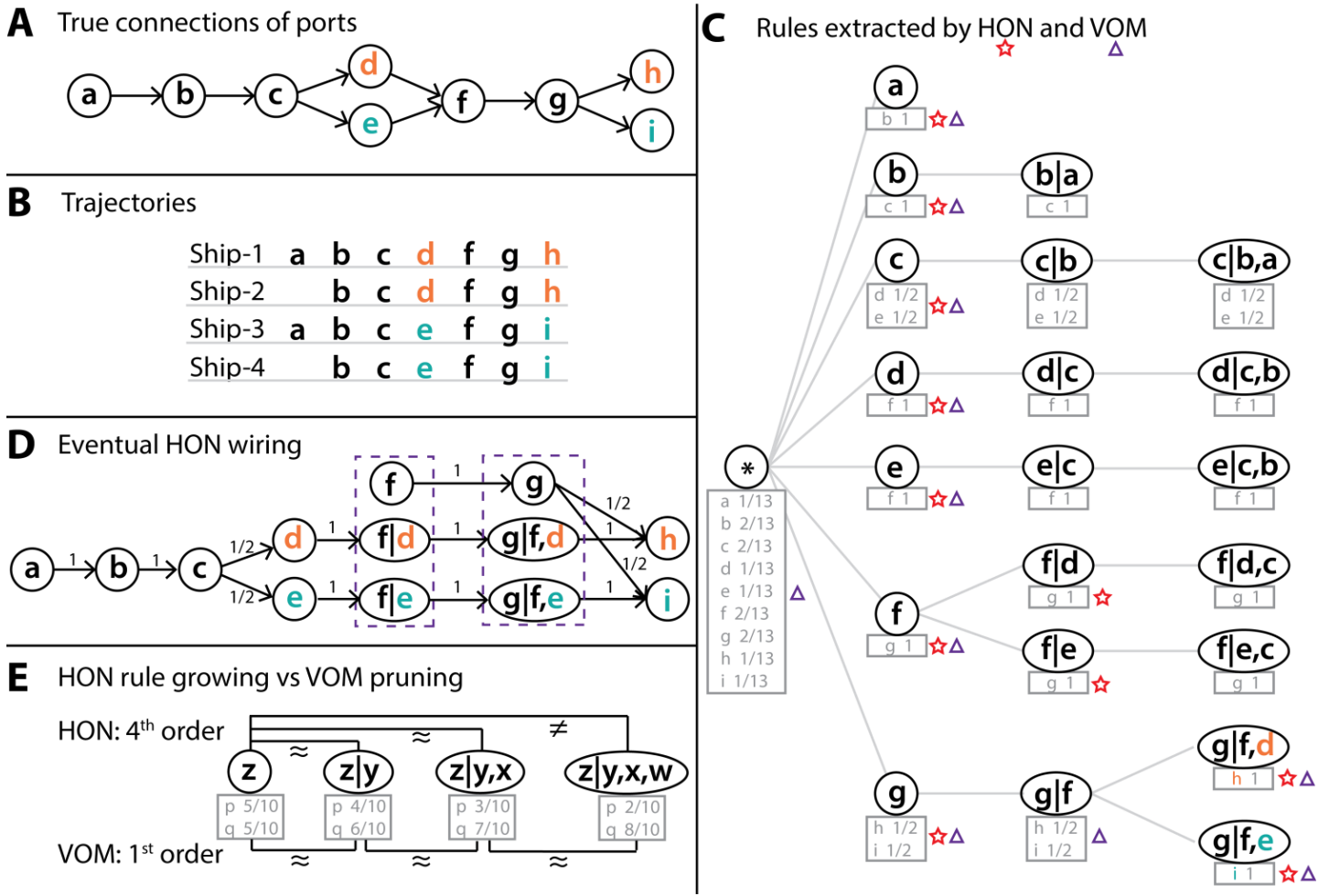

**Fig. S4. Comparison between the HON and the VOM model.** (**A**) In the context of global shipping, the true connection of ports. *f* and *g* are at the two sides of a canal. Ships coming from *d* will go to *h*, and coming from *e* will go to *i*. (**B**) Possible trajectories of ships. (**C**) Comparing the nodes retained by HON and VOM. VOM prunes nodes that are necessary for network representation while retaining nodes that are not necessary. (**D**) The eventual HON representation captures higher-order dependencies while retaining all first-order information. (**E**) HON "grows" rules from the first order, while VOM prunes rules from the highest order.

**Table S1. Changes of PageRank scores by using HON instead of a first-order network.** For the clickstream data. Top 15 gainers / losers of PageRanks scores are listed.

| Pages that gain PageRank scores | ΔPageRank |
|---|---|
| South Bend Tribune - Home. | +0.0119 |
| Hagerstown News / obituaries - Front. | +0.0115 |
| South Bend Tribune - Obits - 3rd Party. | +0.0112 |
| South Bend Tribune / sports / notredame - Front. | +0.0102 |
| Aberdeen News / news / obituaries - Front. | +0.0077 |
| WDBJ7 - Home. | +0.0075 |
| KY3 / weather - Front. | +0.0075 |
| Hagerstown News - Home. | +0.0072 |
| Daily American / lifestyle / obituaries - Front. | +0.0054 |
| WDBJ7 / weather / closings - Front. | +0.0048 |
| WSBT TV / weather - Front. | +0.0041 |
| Daily American - Home. | +0.0036 |
| WDBJ7 / weather / radar - Front. | +0.0036 |
| WDBJ7 / weather / 7-day-planner - Front. | +0.0031 |
| WDBJ7 / weather - Front. | +0.0019 |
| **Pages that lose PageRank scores** | **ΔPageRank** |
| KTUU - Home. | −0.0057 |
| KWCH - Home. | −0.0031 |
| Imperial Valley Press - Home. | −0.0011 |
| Hagerstown News / sports - Front. | −0.0005 |
| Imperial Valley Press / classifieds / topjobs - Front. | −0.0004 |
| Gaylord - Home. | −0.0004 |
| WDBJ7 / weather / web-cams - Front. | −0.0004 |
| KTUU / about / meetnewsteam - Front. | −0.0003 |
| Smithsburg man faces more charges following salvag - Hagerstown News / news - story. | −0.0003 |
| KWCH / about / station / newsteam - Front. | −0.0003 |
| South Bend Tribune / sports / highschoolsports - Front. | −0.0003 |
| Hagerstown News / opinion - Front. | −0.0002 |
| WDBJ7 / news / anchors-reporters - Front. | −0.0002 |
| Petoskey News / news / obituaries - Front. | −0.0002 |
| KWCH / news - Front. | −0.0002 |

**Table S2**. **Comparison of the number of rules extracted from HON and VOM**. Both algorithms are applied on the same global shipping data set, with the same parameters.

|  | HON | VOM | In HON but not in VOM | In VOM but not in HON |
|---|---|---|---|---|
| 0$^{th}$ order | 0 | 3,029 | 0 | 3029 |
| 1$^{st}$ order | 31,028 | 31,028 | 0 | 0 |
| 2$^{nd}$ order | 32,960 | 35,288 | 427 | 2,755 |
| 3$^{rd}$ order | 15,642 | 21,536 | 550 | 6,444 |
| 4$^{th}$ order | 4,632 | 8,973 | 302 | 4,643 |
| 5$^{th}$ order | 763 | 2,084 | 23 | 1,344 |
| Total | 85,025 | 101,938 | 1,302 | 18,215 |

# Algorithm 1. The rule extraction algorithm.

---

**Algorithm 1** Rule extraction. Given the sequential data $T$, outputs higher-order dependency rules $R$.
*Parameters*: $\mathcal{O}$: MaxOrder, $S$: MinSupport

---

1: *global* counter $C \leftarrow \emptyset$
2: *global* nested dictionary $D \leftarrow \emptyset$
3: *global* nested dictionary $R \leftarrow \emptyset$
4:
5: **function** EXTRACTRULES($T$)
6:     BUILDOBSERVATIONS($T$)
7:     BUILDDISTRIBUTIONS()
8:     GENERATEALLRULES()
9:     **return** $R$
10:
11: **function** BUILDOBSERVATIONS($T$)
12:     **for** $t$ in $T$ **do**
13:         **for** $o$ from 2 to $\mathcal{O}$ **do**
14:             $SS \leftarrow$ ExtractSubSequences($t, o$)
15:             **for** $s$ in $SS$ **do**
16:                 $Target \leftarrow$ LastElement($s$)
17:                 $Source \leftarrow$ AllButLastElement($s$)
18:                 IncreaseCounter($C[Source][Target]$)
19:
20: **function** BUILDDISTRIBUTIONS()
21:     **for** $Source$ in $C$ **do**
22:         **for** $Target$ in $C[Source]$ **do**
23:             **if** $C[Source][Target] < S$ **then**
24:                 Remove($C[Source][Target]$)
25:         **for** $Target$ in $C[Source]$ **do**
26:             $D[Source][Target] \leftarrow C[Source][Target]$ / Sum($C[Source][*]$)
27:
28: **function** GENERATEALLRULES()
29:     **for** $Source$ in $D$ **do**
30:         **if** length($Source$)= 1 **then**
31:             ADDTORULES($Source$)
32:             EXTENDRULE($Source, Source, 1$)
33: *(To be continued on the next page.)*

**Algorithm 1** *(continued)*

---

34: **function** EXTENDRULE(*Valid, Curr, Order*)
35:     **if** $Order \geq \mathcal{O}$ **then**
36:         ADDTORULES(*Valid*)
37:     **else**
38:         $Distr \leftarrow D[Valid]$
39:         $NewOrder \leftarrow Order + 1$
40:         $Extended \leftarrow$ all $Source$ satisfying length($Source$)= $NewOrder$ and end with $Curr$
41:         **if** $Extended = \emptyset$ **then**
42:             ADDTORULES(*Valid*)
43:         **else**
44:             **for** $ExtSource$ in $Extended$ **do**
45:                 $ExtDistr \leftarrow D[Extended]$
46:                 **if** KL-Divergence($Distr, ExtDistr$)$> \frac{NewOrder}{\log_2(\text{Sum}(C[ExtSource][*]))}$ **then**
47:                     EXTENDRULE($ExtSource, ExtSource, NewOrder$)
48:                 **else**
49:                     EXTENDRULE($Valid, ExtSource, NewOrder$)
50:
51: **function** ADDTORULES(*Source*)
52:     **if** length($Source$) $> 0$ **then**
53:         $R[Source] \leftarrow C[Source]$
54:         $PrevSource \leftarrow$ AllButLastElement($Source$)
55:         ADDTORULES($PrevSource$)

# Algorithm 2. The network wiring algorithm.

**Algorithm 2** Network wiring. Given the higher-order dependency rules $R$, convert the rules of variable orders into edges, perform rewiring, and output the graph $G$ as HON.

```
 1: function BUILDNETWORK(R)
 2:     global nested dictionary G ← ∅
 3:     R ← Sort(R, ascending, by length of key Source)
 4:     for r in R do
 5:         G.add(r)
 6:         if length(r.Source) > 1 then
 7:             REWIRE(r)
 8:     REWIRETAILS()
 9:     return G
10:
11: function REWIRE(r)
12:     PrevSource ← AllButLastElement(r.Source)
13:     PrevTarget ← LastElement(r.Source)
14:     if (edge: (Source: PrevSource, Target: r.Source)) not in G then
15:         G.add(edge: (Source: PrevSource, Target: r.Source, Weight: r.Probability))
16:         G.remove(edge: (Source: PrevSource, Target: PrevTarget))
17:
18: function REWIRETAILS()
19:     ToAdd ← ∅; ToRemove ← ∅
20:     for r in R do
21:         if length(r.Target) = 1 then
22:             NewTarget ← concatenate(Source, Target)
23:             while length(NewTarget) > 1 do
24:                 if NewTarget in (all Sources of R) then
25:                     ToAdd.add(edge:(Source: r.Source, Target: NewTarget, Weight: r.Probability))
26:                     ToRemove.add(edge:(Source: r.Source, Target: r.Target))
27:                     Break
28:                 else
29:                     PopFirstElement(NewTarget)
30:     G ← (G ∪ ToAdd)\ToRemove
```